\documentclass[ preprint]{aastex}

\usepackage{emulateapj5}
\usepackage{subfigmat}
\usepackage{subfigure}
\usepackage{lineno}
\usepackage{color}

\usepackage{apjfonts}

\slugcomment{}

\shorttitle{Diffuse X-ray Emission from the Loop I Arc}
\shortauthors{Akita et al. 2018}

\begin{document}

\title{Diffuse X-ray emission from the northern arc of Loop I observed with $Suzaku$  \\}


\author{Masahiro Akita\altaffilmark{1}, Jun Kataoka\altaffilmark{1}, Makoto Arimoto\altaffilmark{1,2}, Yoshiaki Sofue\altaffilmark{3}, Tomonori Totani\altaffilmark{3,4}, Yoshiyuki Inoue\altaffilmark{5}, Shinya Nakashima\altaffilmark{5}}
\email{hoppe1031@akane.waseda.jp, kataoka.jun@waseda.jp}


\affil{\altaffilmark{1}Graduated School of Advanced Science and Engineering, Waseda University, 3-4-1, Ohkubo, Shinjuku, Tokyo, 169-8555, Japan}
\altaffiltext{2}{Faculty of Mathematics and Physics, Institute of Science and Engineering, Kanazawa University, Kakuma, Kanazawa, Ishikawa 920-1192, Japan}
\altaffiltext{3}{Department of Astronomy, The University of Tokyo, Bunkyo-ku, Tokyo 113-0033, Japan}
\altaffiltext{4}{Research Center for the Earth Universe (RESCEU), the University of Tokyo, Hongo, Tokyo 113-0033, Japan}
\altaffiltext{5}{RIKEN, 2-1 Hirosawa, Wako, Saitama, 351-0198, Japan}

\begin{abstract}
After discovery of the Fermi bubbles, giant structures observed in radio to X-rays have been discussed as possible evidence of past activities in the Galactic Center (GC).  We report here on the analysis of $Suzaku$ data pointing around the Loop I arc. The diffuse X-ray emission was well represented by the three-component model: (1) an unabsorbed thermal plasma with $kT$ $\simeq$ 0.1~keV either from the Local Hot Bubble (LHB) and/or solar wind charge exchange (SWCX),
  (2) an absorbed thermal plasma regarded as a contribution from the Loop I and the Galactic halo (GH), and (3) an absorbed power-law component representing the cosmic X-ray background. The temperature of the absorbed thermal plasma was clustered in a range of 0.30 $\pm0.02$ keV along Loop I (``ON'' regions), whereas the temperature was about 20 \% lower in the cavity adjacent to the bubbles and Loop I (``OFF'' regions) with 0.24$\pm0.03$ keV.  The emission measure (EM) varied along the Galactic latitude, and was well correlated with the count rate variation as measured with the $ROSAT$ in 0.75~keV band. Although the amount of neutral gas was not conclusive to constrain on the distance to Loop I, the observed EM values rule out a hypothesis that the structure is close to the Sun; we argue that the Loop I is a distant, kpc structure of the shock-heated GH gas.
We discuss the origin of apparent mismatch in the morphologies of the Fermi bubbles and the Loop I arc, suggesting a two-step explosion process in the GC.

\end{abstract}





\section{INTRODUCTION}
Loop I is the largest loop spanning 100$^{\circ}$ on the sky, first detected in the radio wavelength over 50 years ago (Berkhuijsen et al. 1971). A similar arc is also clearly seen in the $ROSAT$ all sky map in X-rays (Snowden et al.1995) and even in the gamma-ray skies (Ackermann et al. 2014). The brightest arm of Loop I is known as the North Polar Spur (NPS). Despite a long history of observations from radio to gamma rays, the origin and distance of these structures are unknown, yet completely different ideas are being proposed.

The most common scenario assumes the NPS/Loop I is the structure close to the Sun, possibly created by a stellar wind from the Scorpio-Centaurus OB association at a distance of 170 pc  (Egger \& Aschenbach 1995) or by a nearby supernova remnant  (Berkhuijsen et al.1971).  The brightest parts of the NPS were recently observed with $XMM$-$Newton$ for detailed spectral studies. Willingale et al. (2003) reported that the X-ray spectra of the NPS are well represented by a thin thermal plasma emission with its temperature
$kT$ $\simeq$ 0.26 keV. Miller et al. (2008) confirmed the presence of $kT$ $\simeq 0.3$~keV plasma with various emissions lines from O, Ne, Mg, and Fe by using the $Suzaku$ satellite (Mitsuda et al. 2007). In these papers, the authors assumed the NPS is a local, hot structure unrelated to the Galactic halo (GH) gas.
A hint of a large amount of neutral matter absorbing the X-ray emission was obtained, but was attributed to the cold gas distribution in the wall located at several tens of pc from the Sun.

An alternative scenario assumes that the NPS/Loop I is a distant, kpc-scale structure in the GH. Already in the 1970's, the NPS/Loop I was suggested to be a bright remnant of the bipolar hyper-shell that was created by an explosion at the Galactic center (GC) $\simeq 15$~M years ago (Sofue et al. 1977, 2000; also Kataoka et al. 2018 for a recent review). This idea was, however, left behind over 40 years until the discovery of the Fermi bubbles by $Fermi$-LAT (Su et al. 2010; Ackermann et al. 2014). The Fermi bubbles are giant gamma-ray structures extending for about 8 kpc above/below the Galactic plane. In the microwave band, the existence of similar giant bubbles, known as ``$WMAP$ haze'', was recently confirmed by $Planck$ observations (Planck Collaboration et al. 2013). In addition, polarized giant radio lobes emanating from the GC seems to be closely connected to the Fermi bubbles (Carrettti et al. 2013). Moreover, the northeastern edge of bubbles seems to have a close contact with the X-ray bright NPS (Su et al. 2010).

\begin{figure*}[t]
\centering
\includegraphics[angle=0,scale=.53]{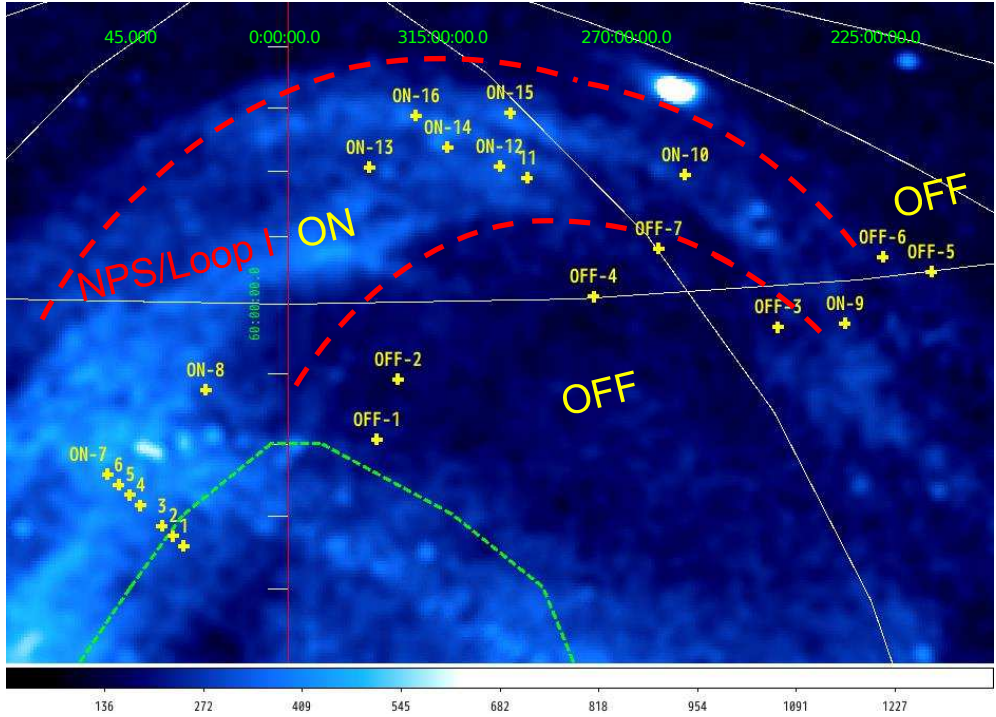}
\vspace{1truemm}
\caption{ \small ROSAT image of the Loop I arc region at 0.75 keV band  in Galactic coordinates, where the white lines represents the same Galactic longitude given by green labels. Yellow crosses indicate the pointing center of the Suzaku archival observations.  The green dotted line is the northern edge of the Fermi bubbles as suggested by Su et al. 2010. The color bar at the bottom indicates the X-ray intensity in units of $\rm{ 10^{-6} cts }$  $\rm{s^{-1} arcmin^{-2}}$. The red dotted lines show the approximate boundaries the ON and OFF regions in the NPS (see Section 2) to guide the eyes.}
\end{figure*}

In this context, Kataoka et al. 2013, 2015 (Paper I and I\hspace{-.1em}I respectively) and Tahara et al. 2015 (Paper I\hspace{-.1em}I\hspace{-.1em}I) conducted deep $Suzaku$ observations of the northeastern and southern edges of the Fermi bubbles. Together with subsequent analysis of more than 100 archival data obtained with $Suzaku$ and the $Swift$ satellites, they confirmed that not only the NPS but the whole extent of the bubble is surrounded by thin thermal plasma of $kT$ $\simeq$ 0.30$\pm$0.07~keV obscured by the neutral hydrogen column density close to the Galactic value $N_{\rm H,Gal}$ (Dickey and Lockman 1990).
They also suggested that a weak shock driven by the bubbles' expansion mildly heated the GH gas from $kT$$\simeq$ 0.2~keV to 0.3~keV with a corresponding shock velocity of $\simeq$ 300 km ${\rm s^{-1}}$.
In fact, recent X-ray tomographic studies further revealed that the NPS is not a nearby structure. Sofue (2015) estimated a distance to the NPS by using  the $ROSAT$ archival X-ray data  around the Aquila Rift region  and estimated the lower limit of the distance as $1.01\pm 0.25$ kpc. Similarly, Lallement et al. (2016) also estimated a distance from 300 pc to 4 kpc. Further, Gu et al. (2016) reanalyzed the NPS data  in Willingale et al. (2003) and Miller et al. (2008) and found that an additional absorption component is required to explain the observed X-ray data.

However, observational properties of the other part of Loop~I are far from being
understood, or even have been undiscussed since the 1970s. Unlike the apparent interaction 
between the NPS and the bubbles (Paper~I), a ``cavity'' exists between Loop I and the northernmost bubble edge at $b$ $> 60^{\circ}$ (Kataoka et al. 2018).  Moreover, Loop I is highly asymmetric above and below the GC, which is inconsistent with the symmetric morphology of Fermi bubbles. Very recently, gamma-ray spectra associated with Loop I have been widely discussed, but are much softer than the bubbles' gamma-ray emissions (Ackermann et al. 2014, Acero et al. 2016; Kataoka et al. 2018). Thus, there is no obvious link between the Fermi bubbles and the Loop I arc in general.

In this paper, we present the first spectral analysis of the northern Loop I arc that have not been reported so far, on the basis of archival data observed with  $Suzaku$ X-ray Imaging Spectrometer (XIS, Koyama et al. 2007). Then we try to understand the origin of the X-ray emission associated with Loop I in conjunction with the NPS and other X-ray structures discussed in Paper I -- III.
In $\S$ 2, we describe the $Suzaku$ observations and data reduction. We describe the details of the spectral analy-
\vspace{0.3cm}
\centerline{{\vbox{\epsfxsize=250pt \epsfysize=210pt \epsfbox{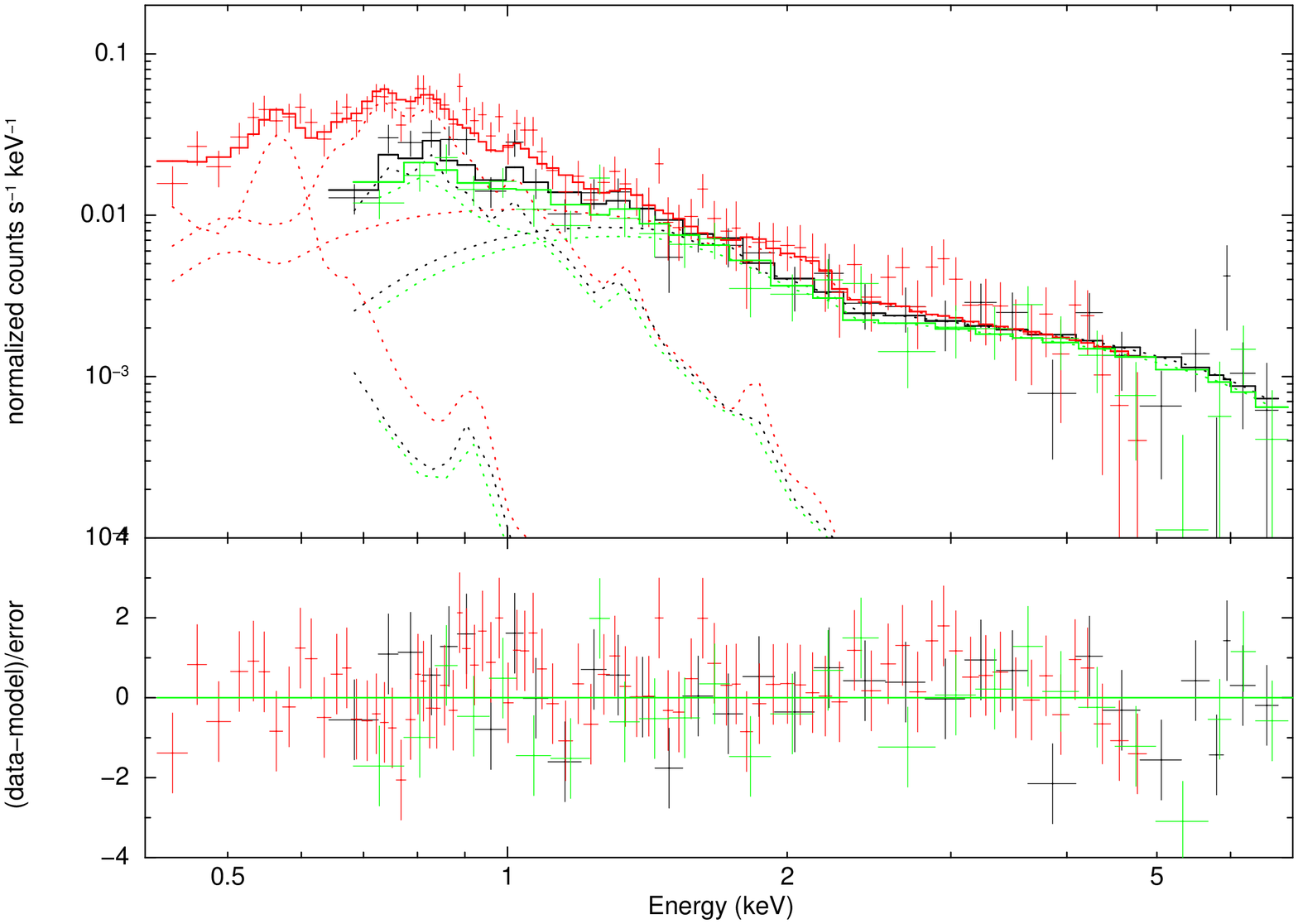}}}}
\vspace{5truemm}
\figcaption{\small  The typical spectra and the best-fit model curve of the diffuse X-ray emission for the NPS/LOOP1 region (ON-14). The $black$ crosses/line show the XIS0 data/model-curve, the $red$ shows XIS1, and the $green$  shows XIS3, respectively. The $dotted$-$curve$ shows the contribution of each model component (see text for more details).}
\vspace{0.5cm} \noindent
sis and the obtained results in $\S$ 3.  We summarize our results and discuss the physical origin of the northern Loop I in $\S$ 4. All statistical errors in texts and tables are $1\sigma $ unless otherwise stated.

\begin{table*}[htb]
\begin{center}
\caption{Suzaku observations.\label{AG_param}}
\begin{tabular}{cccccccc}
\hline\hline
Name & ObsID & RA\tablenotemark{a}    & DEC\tablenotemark{b}  & l\tablenotemark{c}      &       b\tablenotemark{d} &  Exposure Time\tablenotemark{e} \\
         &          &  (deg) & (deg) & (deg) & (deg) &  (ksec) \\
\hline
OFF-1 & 508007010 & 221.750 & -1.315 & 351.952 & 50.223 & 20.7 \\ 
OFF-2 & 807062010 & 217.761 & 0.788 & 349.304 & 54.434 & 15.3 \\
OFF-3 & 707001010 & 196.053 & -5.559 & 308.801 & 57.167 & 65.3 \\ 
OFF-4 & 702067010 & 204.195 & -0.819 & 326.078 & 59.999 & 11.8 \\ 
OFF-5 & 704006010 & 184.787 & -1.811 & 286.613 & 60.033 & 13.3 \\ 
OFF-6 & 701001010 & 186.707 & -0.937 & 290.032 & 61.315 & 36.7 \\ 
OFF-7 & 705011010 & 198.188 & 0.850 & 314.829 & 63.228 & 16.6 \\

\hline
ON-1 & 507008010 & 234.713 & 2.194 & 8.3271 & 42.834 & 12.0 \\ 
ON-2 & 507007010 & 234.551 & 3.168 & 9.2841 & 43.533 & 17.0 \\ 
ON-3 & 507002010 & 234.405 & 4.124 & 10.256 & 44.200 & 19.0 \\ 
ON-4 & 507003010 & 234.034 & 6.090 & 12.270 & 45.602 & 16.7 \\ 
ON-5 & 507004010 & 233.833 & 7.080 & 13.313 & 46.305 & 8.8 \\ 
ON-6 & 507005010 & 233.623 & 8.072 & 14.377 & 47.007 & 13.2 \\ 
ON-7 & 507006010 & 233.400 & 9.071 & 15.472 & 47.712 & 17.5 \\ 
ON-8 & 802038010 & 225.629 & 8.293 & 7.819 & 53.735 & 19.1 \\ 
ON-9 & 805041010 & 192.151 & -5.791 & 301.636 & 57.074 & 73.9 \\ 
ON-10 & 802039010 & 192.513 & 5.456 & 301.998 & 68.325 & 25.4 \\ 
ON-11 & 509062010 & 200.607 & 7.382 & 324.764 & 68.930 & 14.3 \\ 
ON-12 & 509059010 & 201.171 & 8.665 & 327.544 & 69.932 & 18.2 \\ 
ON-13 & 702053010 & 206.913 & 12.350 & 347.389 & 70.203 & 40.3 \\
ON-14 & 804048010 & 202.357 & 11.020 & 333.767 & 71.580 & 35.8 \\ 
ON-15 & 701080010 & 197.244 & 11.578 & 318.614 & 73.913 & 29.6 \\ 
ON-16 & 805074010 & 201.743 & 13.573 & 336.170 & 74.105 & 11.3 \\
\hline
\end{tabular}
\tablenotetext{a}{R.A. of the $Suzaku$ pointing center in J2000 equinox }
\tablenotetext{b}{Decl. of the $Suzaku$ pointing center in J2000 equinox}
\tablenotetext{c}{Galactic longitude of the $Suzaku$ pointing center}
\tablenotetext{d}{Galactic latitude of the $Suzaku$ pointing center}
\tablenotetext{e}{Exposure time of good time interval after the data reduction described in $\S$~2.}
\end{center}
\end{table*}

\section{OBSERVATIONS and DATA REDUCTION}\label{Sec:Observation}
We analyzed the archival data of the $Suzaku$ XIS pointed around the NPS/Loop I regions. We excluded pointings that contains either extended X-ray sources such as galaxy clusters, or too bright X-ray sources whose tailed emission in the point spread function (PSF) may affect the analysis results in the same field-of-view (FOV). As a result, 
23 data were selected as listed in Table.1. Total exposure amounts to 552~ks. The pointing centers of each observation are shown in Figure.1. 
For convenience, all the pointings are classified into `ON'' and ``OFF'' regions on the basis of the  $ROSAT$ all-sky map as measured at 0.75~keV (R34) band. The pointings situated on the northern Loop I arc are denoted ON-1 to 16, where ON-1 to 7 denote exactly the same pointings as already reported in Paper I. Relatively faint regions adjacent to the Fermi bubbles and Loop I, but  involved in Loop IV (Large et al. 1966, Berkhuijsen et al. 1971), are denoted OFF-1 to OFF-7. The analysis results of OFF-1 and OFF-2 are already described  in Paper I\hspace{-.1em}I\hspace{-.1em}I.

We extracted the XIS data from XIS~0, 1, 3, as XIS~2 has been unusable since 2006.
Note that XIS~0 and 3 are front-illuminated CCD (FI CCD), whereas XIS~1 is back-illuminated CCD (BI CCD)  that has better sensitivity than the FI-CCDs below 1~keV.
In  the reduction procedure and the analysis of the all $Suzaku$ data, we used \textsc{headas} software version 6.17, \textsc{xspec} version 12.9.0, and the calibration databases (CALDB) released on April 2016. First, we combined the data edited with different observation modes (3$\times$3, 5 $\times$5) using \textsc{xselect}. Then we ran the \textsc{sisclean} command for each data to remove hot pixels and flickering pixels. We applied a new method using \textsc{noisy pixel map} \footnote{ http:\slash{}\slash{}www.astro.isas.ac.jp\slash{}suzaku\slash{}analysis\slash{}xis\slash{}nxb\_new2\slash } to remove flicking pixels more precisely. After the cleaning, we processed the data reduction with the following criteria: (1) a cut off rigidity (COR) larger than 6 GeV, (2) the elevation angle from the day and night Earth less than 20 degree, and (3) exclusion of the data during passage through the South Atlantic Anomaly and after 436 s. In addition, (4) we excluded the time interval during the proton flux in the Solar wind larger than $4.0 \times 10^8 {\rm cm^{-2} s^{-1}}$to reduce the geo-coronal Solar Wind Charge Exchange (SWCX)  referred to by Yoshino et al. (2009). To calculate the proton flux, we used the data of the $ACE$ and the $WIND$ satellite \footnote{ https:\slash{}\slash{}omniweb.gsfc.nasa.gov\slash{}form\slash{}dx1.html }.

\section{Analysis and Results}\label{Sec:Analysis}

 \subsection{Extraction the diffuse emission}\label{Sec:Extraction the diffuse emission}
Before extracting the X-ray spectrum of diffuse emission, we constructed the XIS images in 0.4$-$10 keV and 0.4$-$2.0 keV bands for each observation. Two FI CCD images (XIS0 and XIS3) were merged to increase photon statistics.  We applied the source detection algorithm \textsc{detect} separately in the \textsc{ximage} for the XIS0+3 and the XIS1 images in the two energy bands, and then concatenated the results. We removed the contribution from possible point sources detected above 3$\sigma$ in the same FOV by setting the 2' radius centered on the detected sources. We then extracted the spectra from the remaining source region over the CCD chip. In the spectral fitting, we generated RMFs (Redistribution Matrix Files) and ARFs (Auxillary Response Files) with \textsc{xisrmfgen} and \textsc{xissimarfgen} (Ishisaki et al. 2007) and the non-X-ray background spectra from the night Earth observation data with \textsc{xisnxbgen} (Tawa et al. 2008).

 \subsection{Spectral analysis}\label{Sec:SpectralAnalysis}

For the spectral analysis, we used 0.4$-$5.0~keV data for the XIS~1 and 0.6$-$7.0~keV data for the XIS~0, 3, except the analysis of OFF-3 and ON-12. 
The spectra of OFF-3 exhibited a large deviation owing to the Si-K edge structure, and thus we excluded the data between 1.7 and 2.0 keV. 
For the spectra of ON-12, there remained a strong line due to instrumental background (Al ${\rm K_\alpha}$, 1.5 keV) after subtracting the NXB spectra, and therefore we excluded the data between 1.4 and 1.6 keV.

We fitted all the spectra using \textsc{xspec} with a model consisting of three plasma components as applied in previous works (e.g. Yoshino et al. 2009, Paper I, I\hspace{-.1em}I.); (1) an unabsorbed thermal plasma with $kT$ $\simeq$ 0.1~keV either from the Local Hot Bubble (LHB) and/or the solar wind charge exchange (SWCX), (2) an absorbed thermal plasma regarded as a contribution from Loop I and GH, and (3) an absorbed power-law component representing the cosmic X-ray
background (CXB). For the first and second thermal components, we adopted \textsc{apec} model, which represents an emission spectrum from collisionally-ionized diffused gas calculated from the AtomDB atomic database. For the LHB/SWCX emission, we assumed the metal abundance Z= ${\rm Z_\odot}$ (Smith et al. 2007, Yoshino et al. 2009, Willingale et al. 2003, Miller et al. 2008). For Loop I, we fixed the metal abundance to be ${\rm Z=0.2~Z_{\odot}}$, an average value for the northern east part of the NPS (Appendix-B of Paper I). 
For the third CXB,  we fixed the photon index $\Gamma =1.41$ as determined by Kushino et al. (2002).

We also note that the spectra of ON-12 and 13 showed a significant residual around 0.52keV, most likely due to a contamination by an ${\rm O_I}$ fluorescent line (center energy E = 0.525 keV) produced in the Earth's atmosphere. 
Sekiya et al. (2014) showed that the contamination can be reduced by filtering of the elevation angle from the day Earth limb (\textsc{dye\_elv}). Although we applied various filters with \textsc{dye\_elv = 60} or more, the strong ${\rm O_I}$ line remained, even though the good time interval (GTI) decreased significantly. We thus added a gaussian-like line function by fixing \textsc{dye\_elv} to 60 deg ON-12 and ON-13. 
In addition, we found that the spectrum of ON-15 has a discrepancy at approximately 0.5 keV. When the metal abundance of nitrogen was left free, the discrepancy became small, the best-fit value is N/O $= 5.2_{-1.1}^{+1.2}$ and the $\chi^2$ improved from 235.65 (191 dof) to 222.30 (190 dof) with a statistical significance at the 3 $\sigma$ confidence level. Such a large enhancement in nitrogen abundance was also reported in Miller et al. (2008), but Gu et al. (2016) claimed that the large enhance can be decreased by adding an ionized absorption. Although we cannot investigate the line emission ratio and the ionized absorption like Gu et al. (2016) because of poor photon statistics and shorter exposure time, the results will not affect the observed $kT$ and EM discussed in this paper.

\subsection{Distribution of $kT$ and EM}\label{Sec:Results2}

Figure.2 shows an example spectrum of ON-14 as represented by the above three-components  plasma model. 
A characteristic line feature below 0.6 keV is due to the ${\rm O_{VI\hspace{-.1em}I}}$ line (0.574 keV) which is emitted  LHB/SWCX component. In contrast, the ${\rm O_{VIII}}$ line at 0.654 keV is emitted from the GH and/or the Loop I emission component. Above 2 keV the CXB is dominant with no spectral line features. In both the ON and OFF regions, the obtained flux of the CXB component is consistent with the average CXB flux given by Kushino et al. (2002) within a reported fluctuation of CXB of approximately
10$\%$.

The results are summarized in Table.2 and Figure.3.
In the cases where the neutral absorption column density $N_{\rm H}$ was left free
in the spectral fitting, the ratio to the full Galactic column ($N_{\rm H}$/$N_{\rm H,Gal}$) became larger than 1 or unconstrained in most of the regions (see Table.2). Therefore, we fixed the $N_H$ as $N_{\rm H,Gal}$ for all the cases, as has been done in Paper I and II. 
We also note that results are consistent with Paper I for ON-1 to ON-7
within the statistical uncertainties.

\begin{table*}[htb]
\centering
\caption{Results of the spectral analysis.\label{AG_param}}
\small
\begin{tabular}{ccccccccc}
\hline\hline
Name & $N_{H,Gal}$\tablenotemark{a} & $N_{H}/N_{H,Gal}$\tablenotemark{b} & ${\rm EM_{LHB}}$\tablenotemark{c} & ${\rm kT_{GH}}$\tablenotemark{d} &  ${\rm EM_{GH}}$\tablenotemark{e} & CXB  & ${\rm O_{I}}$ &  $\chi^{2}$/d.o.f   \\
         &  ($10^{20}{\rm cm^{-2}}$) &   & ($10^{-2}{\rm cm^{-6}}$pc) & (keV) & ($10^{-2}{\rm cm^{-6}}$pc) & Norm \tablenotemark{f} & Norm\tablenotemark{g}  & \\
\hline
OFF-1 & 4.12 & 1.5 - 9.9 & $1.35_{-0.40}^{+0.40}$  & $0.298_{-0.021}^{+0.028}$ & $4.77_{-0.80}^{+0.80}$ &$0.84_{-0.04}^{+0.04}$ & - &140.04/145 \\
OFF-2 & 3.02 &  $<$ 16.2  & $1.73_{-0.87}^{+0.81}$  & $0.265_{-0.033}^{+0.053}$ & $4.72_{-1.55}^{+2.03}$ &$0.69_{-0.05}^{+0.04}$ & - &120.37/103 \\ 
OFF-3 & 2.45 & 13.4 - 24.0  & $1.98_{-0.42}^{+0.42}$  & $0.241_{-0.017}^{+0.026}$ & $1.77_{-0.40}^{+0.40}$ &$0.84_{-0.04}^{+0.04}$ & - &175.87/190 \\
OFF-4 & 1.98 &  $<$ 5.7  & $2.40_{-0.33}^{+0.33}$  & $0.25_{fixed}$ & $1.21_{-0.20}^{+0.20}$ & $0.66_{-0.06}^{+0.05}$ & - & 96.94/98 \\
OFF-5 & 2.29 & 1.2 - 3.4 & $1.89_{-0.73}^{+0.54}$  & $0.242_{-0.047}^{+0.067}$ & $0.86_{-0.39}^{+0.83}$ &$0.90_{-0.04}^{+0.04}$ & - &119.00/122 \\
OFF-6 & 2.08 & 4.3 - 10.7 & $3.22_{-0.32}^{+0.32}$  & $0.224_{-0.012}^{+0.013}$ & $1.87_{-0.28}^{+0.33}$ &$0.72_{-0.03}^{+0.03}$ & - &134.28/123 \\ 
OFF-7 & 1.93 &  $<$ 11.2 & $1.39_{-0.57}^{+0.53}$  & $0.257_{-0.023}^{+0.030}$ & $1.86_{-0.42}^{+0.52}$ & $0.91_{-0.04}^{+0.04}$ & - &143.86/151 \\
\hline
ON-1 & 5.02 & 3.0 - 7.7 & $3.66_{-0.52}^{+0.52}$  & $0.281_{-0.022}^{+0.015}$ & $4.11_{-0.44}^{+0.81}$ &$0.78_{-0.04}^{+0.04}$ & - &140.71/144 \\ 
ON-2 & 4.76 & 3.0 - 6.7 & $4.96_{-0.51}^{+0.51}$  & $0.306_{-0.011}^{+0.013}$ & $4.26_{-0.35}^{+0.35}$ &$0.62_{-0.03}^{+0.03}$ & - &151.37/148 \\ 
ON-3 & 4.44 & 1.0 - 5.6 & $4.43_{-0.46}^{+0.46}$  & $0.312_{-0.013}^{+0.015}$ & $3.89_{-0.34}^{+0.34}$ &$1.00_{-0.04}^{+0.04}$ & - &172.02/153 \\ 
ON-4 & 4.05 & 0.5 - 3.4 & $4.28_{-0.49}^{+0.49}$  & $0.311_{-0.010}^{+0.011}$ & $5.87_{-0.39}^{+0.39}$ &$0.70_{-0.04}^{+0.04}$ & - &193.56/169 \\
ON-5 & 3.86 & 1.0 - 5.1 & $2.65_{-0.60}^{+0.60}$  & $0.297_{-0.011}^{+0.012}$ & $6.26_{-0.51}^{+0.51}$ &$1.14_{-0.05}^{+0.05}$ & - &173.24/165 \\
ON-6 & 3.83 & 1.4 - 5.4 & $4.67_{-0.64}^{+0.64}$  & $0.328_{-0.014}^{+0.017}$ & $5.30_{-0.47}^{+0.47}$ &$0.95_{-0.04}^{+0.04}$ & - &158.44/138 \\ 
ON-7 & 3.36 & 0.3 - 4.1 & $3.69_{-0.49}^{+0.49}$  & $0.325_{-0.011}^{+0.013}$ & $5.16_{-0.37}^{+0.37}$ &$1.01_{-0.04}^{+0.04}$ & - &166.17/180 \\ 
ON-8 & 2.25 & 3.9 - 12.2 & $2.73_{-0.39}^{+0.39}$  & $0.324_{-0.014}^{+0.017}$ & $3.07_{-0.28}^{+0.28}$ &$0.92_{-0.04}^{+0.04}$ & - &162.63/141 \\ 
ON-9 & 2.14 & $<$ 2.59 & $2.29_{-0.32}^{+0.32}$  & $0.293_{-0.009}^{+0.010}$ & $3.40_{-0.23}^{+0.23}$ &$0.85_{-0.03}^{+0.03}$ & -  &152.72/144 \\
ON-10 & 1.95 & $<$ 13.3 & $2.98_{-0.33}^{+0.33}$  & $0.291_{-0.012}^{+0.014}$ & $2.30_{-0.22}^{+0.22}$ &$0.85_{-0.03}^{+0.03}$ & - &183.27/169 \\ 
ON-11 & 2.06 & 7.5 - 18.9 & $1.11_{-0.59}^{+0.59}$  & $0.305_{-0.015}^{+0.018}$ & $3.25_{-0.34}^{+0.34}$ &$0.76_{-0.04}^{+0.04}$ & $8.0_{-1.8}^{+1.8}$ &172.79/134 \\ 
ON-12 & 1.92 & 8.9 - 17.8 & $2.05_{-0.66}^{+0.66}$  & $0.321_{-0.014}^{+0.017}$ & $3.82_{-0.35}^{+0.35}$ &$0.59_{-0.04}^{+0.04}$ & $11.9_{-2.0}^{+2.0}$ &149.93/145 \\ 
ON-13 & 1.89 & 2.0 - 13.6 & $1.87_{-0.29}^{+0.29}$  & $0.387_{-0.027}^{+0.033}$ & $2.09_{-0.22}^{+0.22}$ &$0.91_{-0.04}^{+004}$ & - &135.47/138 \\
ON-14 & 1.93 & 4.0 - 10.7 & $2.65_{-0.36}^{+0.36}$  & $0.324_{-0.010}^{+0.011}$ & $3.36_{-0.22}^{+0.22}$ &$0.94_{-0.03}^{+0.03}$ & - &162.63/141 \\ 
ON-15 & 2.09 & $>$ 0.4 & $2.17_{-0.40}^{+0.34}$  & $0.269_{-0.008}^{+0.006}$ & $5.17_{-0.27}^{+0.40}$ &$0.75_{-0.03}^{+0.03}$ & - &235.65/191 \\ 
ON-16 & 1.77 & $<$ 13.7 & $2.09_{-1.14}^{+1.14}$  & $0.251_{-0.017}^{+0.023}$ & $4.86_{-0.98}^{+1.06}$ &$0.92_{-0.05}^{+0.05}$ & - &83.53/77 \\ 

\hline
\end{tabular}
\tablenotetext{a}{Galactic values of the absorption column density given in Dickey \& Lockman (1990).} \tablenotetext{b}{The ratio of the absorption column density to the full galactic absorption column density when we left $N_{H}$ free in the spectral fitting. These values are obtained at 3 $\sigma$ confidence level. }
\tablenotetext{c}{Emission measure of the LHB and the SWCX component with the \textsc{apec} model. We fixed the temperature at $kT$ $=$0.1 keV and the metal abundance ${\rm Z = Z_{\odot}}$.}
\tablenotetext{d}{Temeprature of the GH component with the \textsc{apec} model. We fixed the the metal abundance to ${\rm Z = 0.2 Z_{\odot}}$.} 
\tablenotetext{e}{Emission measure of the GH component with the \textsc{apec} model. We fixed  the metal abundance to ${\rm Z = 0.2 Z_{\odot}}$.}
\tablenotetext{f}{The normalization of the CXB with the power-law model for the fixed photon index $\Gamma = 1.41$ in the unit of 5.85 $\times$ ${\rm 10^{-8}erg cm^{-2} s^{-1}sr^{-1}}$ (Kushino et al.2002). }
\tablenotetext{g}{The normalization of the gaussian for the fixed center energy E = 0.525 keV in the unit of L.U. (photons${\rm s^{-1} cm^{-2} str^{-1}}$)}
\end{table*}

Apparently, the temperature ($kT$) of the absorbed thermal emissions is different between ON and OFF regions; it is narrowly concentrated in $kT$ = 0.30 $\pm$ 0.02~keV for the ON regions, and in 0.24 $\pm$ 0.03 keV for OFF regions, except OFF-4. Owing to poor photon statistics, we could not determine the temperature of the GH component of the OFF-4 region, and thus we fixed the temperature to be 0.25 keV for OFF-4.  As shown in Figure.3 ($left$), there is no clear dependence on $kT$ as a function of the Galactic latitudes $b$ for the ON regions. 
The obtained EMs gradually decrease along $b$ and are systematically larger in the ON regions than in the OFF regions by a factor of a few, but increase again in high latitude of  $b$ $>$ 65$^{\circ}$, consistent with the  $ROSAT$ all-sky map in 0.75~keV. Figure.3 (right) shows a scatterplot comparing $kT$ and $EM$, highlighting the difference in the best parameters between ON and OFF regions.

\begin{figure*}[htb]
\centering
\includegraphics[angle=0,scale=.467]{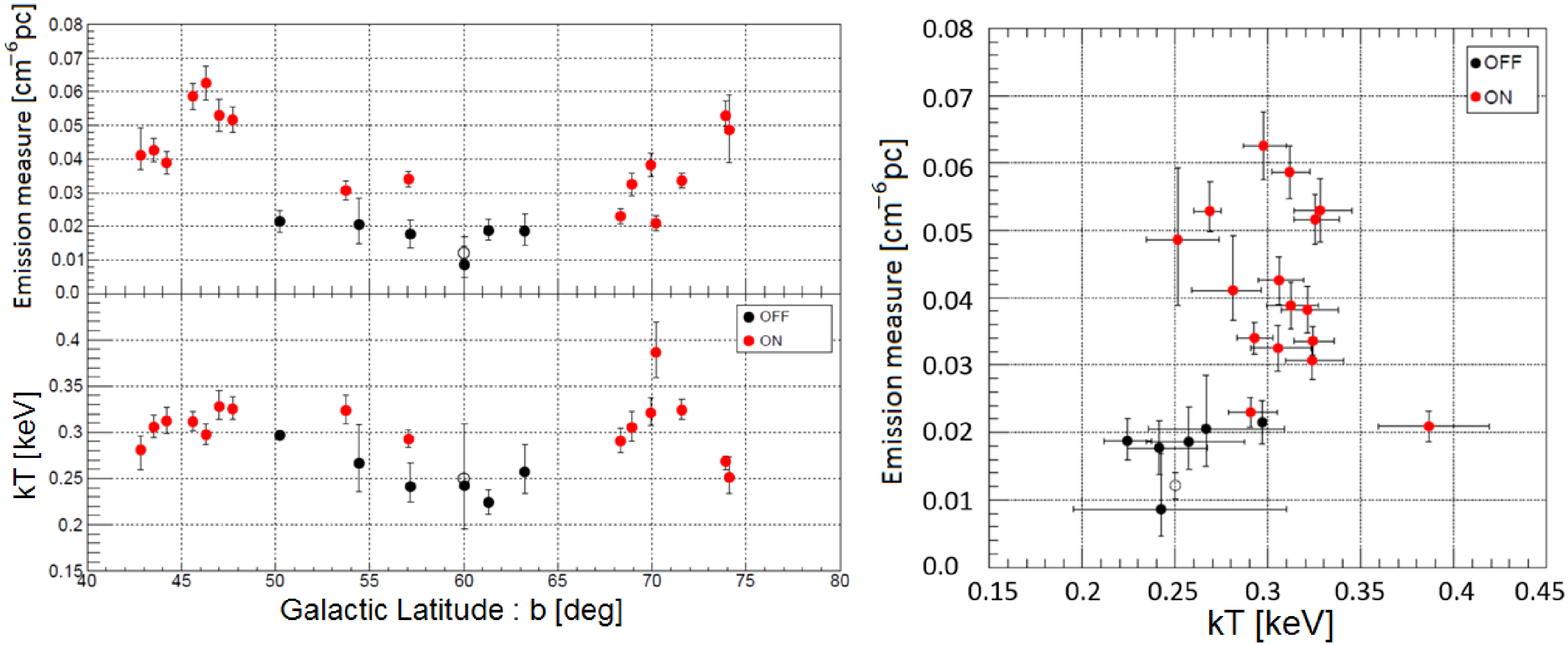}
\vspace{3truemm}
\caption{\small  ($left$) Emission measure ($top$) and temperature ($bottom$) of the absorbed thermal plasma as a function of the Galactic latitude $b$ in the Loop I arc regions (see, Table 2). $Red$ and $black$ points indicate the value of ON/OFF regions and a $black$ $open$ $circle$ indicates the value of OFF-4 when $kT$ is fixed at 0.25~keV. ($right$) A scatterplot showing the emission measure versus temperature of the absorbed thermal plasma component. Same as above for the symbol.}
\end{figure*}

\section{DISCUSSION and CONCLUSION}\label{Sec:Discussion}

\subsection{Loop I; close or distant arc ?}\label{Sec:D_Loop_I}
In the spectral fitting of $Suzaku$ data, we showed that the neutral hydrogen column density $N_{\rm H}$ is close to, or even larger than, the full Galactic column $N_{\rm H,Gal}$ in most cases, although uncertainties are too large to provide any useful constraints on the distance. This fact itself may indicate that Loop I is a distant structure in the GH, as already discussed in Paper~I, but here we try to provide independent estimate of the distance by using the EM as measured for  Loop I. 
In the local scenario, the radio emission of the Loop I is thought to be associated with the high-latitude HI gas, whose distance is constrained to 98 $\pm$ 6 pc in $b$ $\leq$ 70$^{\circ}$ and 95 -- 157~pc in 55$^{\circ}$ $\leq$ $b$ $\leq$ 70$^{\circ}$  (Puspitarini \& Lallement 2012). Similarly, Loop I is located at a distance of 170~pc from the Sun by Egger \& Aschenbach (1995). Assuming the pressure equilibrium between Loop I and the LHB, the electron number density can be estimated as $n_{e}$ = 1.6 $\times$ $10^{-3}$ ${\rm cm^{-3}}$ if there is no large variation in the magnetic field between these structures (Urshino et al. (2015)). Assuming that the density of the Loop I shell is constant, we can estimate the path length $d$ in which the observed EM is obtained as $d$ $\simeq$ EM\slash{}${n^{2}_{e}}$.  
Because the EM measured in the ON regions ranges is in the range (2.09$-$6.26)$\times$10$^{-2}$ ${\rm cm^{-6}}$ pc, we obrtain $d$ $\simeq$ 8.2--24.5~kpc. Instead, if we adopt the electron number density of the LHB, $n_{e}$ = 4.7 $\times$ $10^{-3}$ ${\rm cm^{-3}}$ (Snowden et al. 2014), $d$ would be 0.9--2.8 kpc. In both cases, the corresponding path length is too long and thus is inconsistent with the assumption that Loop I is located at a few hundreds pc from the Sun. Again, we thus argue that the Loop I/NPS are distant structures in the GH. 

\subsection{Shocked vs unshocked halo gas}\label{Sec:OFF_spectra}
In Paper I, II and III, we showed that the absorbed thermal plasma with $kT$ $\simeq$ 0.3~keV is observed ubiquitously in the Fermi bubbles and surrounding NPS. We interpreted this emission as weakly shock-heated GH gas. In this paper, we showed that the temperature of Loop I is also close to $kT$ $\simeq$ 0.3~keV. In this context, the temperature of plasma in the cavity (OFF regions), $kT$ $\simeq$ 0.25~keV, is a little lower, but still higher than the canonical value of GH, $kT$ $\simeq$ 0.2~keV (e.g., Yoshino et al. 2009). To highlight the different temperature of the plasma more clearly, Figure.4 compares the EM obtained in this paper and the $ROSAT$ counting rate as measured in the 0.75 keV band. In addition, four EM values measured in the NPS (see, Willingale et al. 2003; Miller et al. 2007) are also plotted as a reference. 
Here, the $ROSAT$ counting rate in the corresponding fields were estimated by using the X-ray Background Tool at NASA/GSFC\footnote{https:\slash{}\slash{}heasarc.gsfc.nasa.gov\slash{}cgi-bin\slash{}Tools\slash{}xraybg\slash{}xraybg.pl} (Snowden et al. 1997) at the same positions observed by the $Suzaku$ and the $XMM$--$Newton$.  To estimate $ROSAT$ counts, we adopted a larger size ($0.6^{\circ}$) of circles than $Suzaku$ FOV (17' $\times$ 17'), considering the larger PSF of the $ROSAT$-PSPC.

A tight correlation is visible with the correlation coefficient $r$ $=$ 0.98, suggesting that any X-ray bright structures including the NPS and Loop I seen in the $ROSAT$ 0.75~keV map are possibly related and have the same physical origin. However, a closer look at the data reveals that the correlation between the NPS ($open$ $red$)/Loop I (ON: $filled$ $red$) and  cavity/OFF regions ($filled$ $black$) shows a slight mismatch, owing to different plasma temperature. This trend is more clearly seen in the bottom panel, in which the residual to the best-fit linear function, determined for the NPS/Loop I region, are extrapolated to the cavity/OFF region. If  $kT$ $\simeq$ 0.3~keV is the characteristic temperature of shock-heated plasma, we may observe the contribution of the unshocked GH gas ($kT$ $\simeq$ 0.2~keV) and the shocked halo gas ($kT$ $\simeq$ 0.3~keV) at the same time everywhere in the sky. In this context, $kT$ $\simeq$ 0.25~keV plasma may be a result of an almost even combination of shocked/unshocked halo gas for relatively faint regions such as in the cavity. The contribution of a disk-like hot gas in the GH is modeled in Sakai et al. (2014) and is  plotted as a $blue$ $line$ in Figure.5 ($upper$) with possible uncertainties in the model.
Note that Sakai et al. (2014) assumed a solar abundance, whereas we fixed it at ${\rm Z=0.2~Z_{\odot}}$. Therefore in Figure. 4 we normalize the model by taking the correction factor of 3.8, that was determined from the analysis of typical GH gas well outside the Fermi bubbles and NPS/Loop I regions.
Figure.5 ($bottom$) shows the ratio of EM in the shocked and unshocked halo gas, where the contribution of the shocked gas is estimated by subtracting the disk-like model from the observed EM. Note that, the ratio of shocked/unshocked gas is almost unity as expected in the cavity/OFF regions, whereas it ranges from 2 to more than 5 in the ON regions, where the emission is dominated by the $kT$ $\simeq$ 0.3~keV plasma.

\subsection{Two Step Explosion Scenario}\label{Sec:OFFs}

Although the shock-heated plasma as observed in the Loop I arc is most probably related to the NPS in origin, only the morphology of the NPS seen in the $ROSAT$ all-sky map aligns well with the northeastern boundary of the bubble. Such close interaction is hardly seen with the northernmost Loop I arc, leaving a huge cavity as revealed in the paper. Moreover, the only weak sign of the NPS-like feature, called the South Polar Spur (SPS), is seen in the south in both radio and X-rays (e.g., Kataoka et al. 2018). Such an asymmetry between the areas north and south of the GC itself is not surprising, and may be explained by a large-scale outflow from the GC. In fact, most shocked shells, such as SNRs and the GC phenomena, as well as extragalactic jets in the Active Galactic Nuclei (AGN), are more or less asymmetric, similar to the NPS and the SPS. An alternative theory is that the GH has a structural as well as a dynamic asymmetry with respect to the Galactic plane, caused by the intergalactic 
\vspace{0.5cm}
\centerline{{\vbox{\epsfxsize=230pt \epsfbox{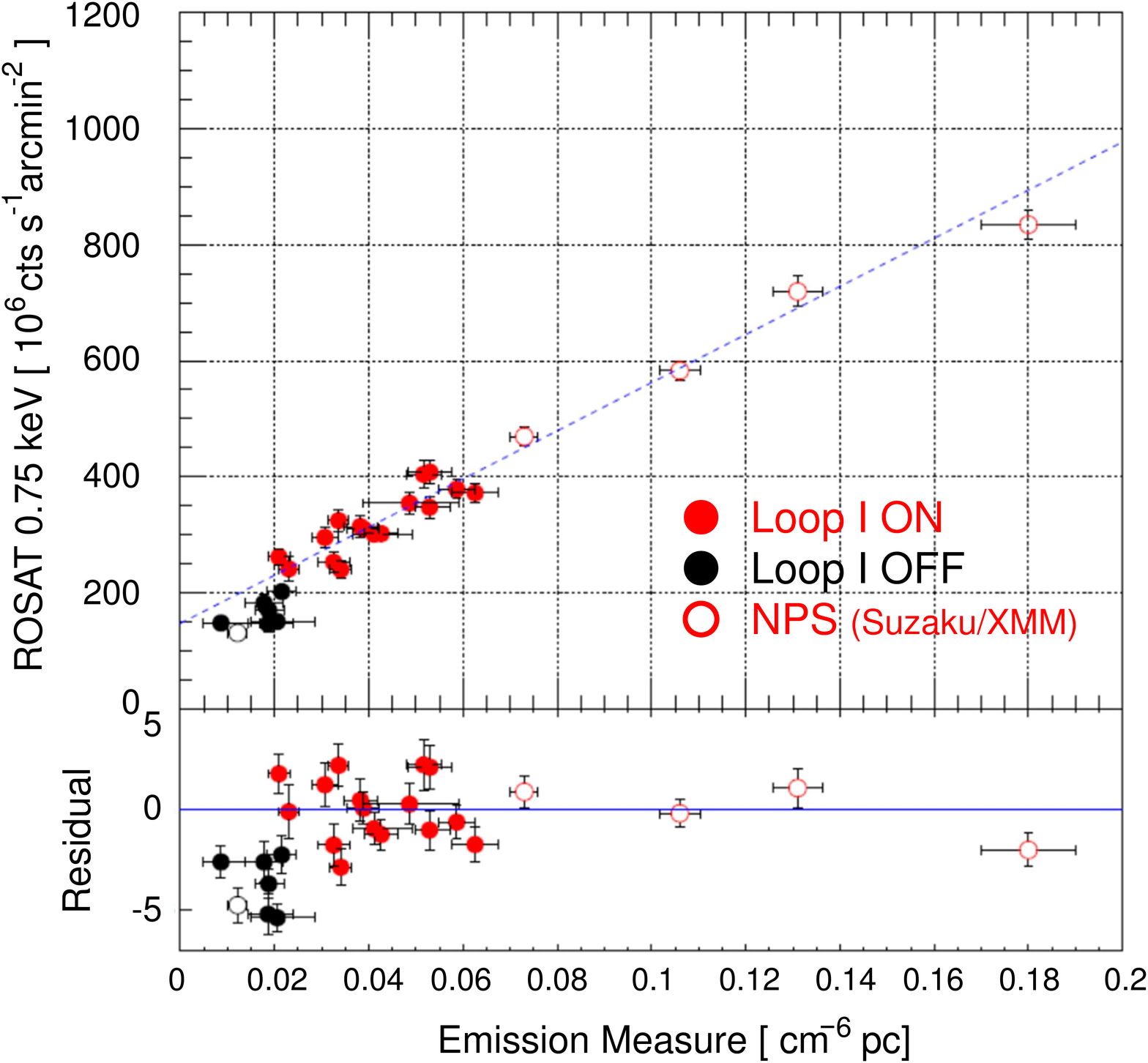}}}}
\figcaption{\small  ($top$) $ROSAT$ counting rate in the 0.75~keV versus the emission measure of the absorbed thermal plasma derived in this paper. $Red$ and $black$ points indicate the value of ON/OFF regions and $red$ $open$ $circles$ indicate the archival EM values of $Suzaku$ and $XMM$-$Newton$ observations for the NPS (Willingale et al. 2003, Miller et al. 2008).
  ($bottom$) residual to the best-fit linear function determined for the NPS and ON data as shown in the $blue$ line in the upper panel. Note that  for the OFF region has a larger residual, due to that plasma having a little lower temperature than in the NPS/ON regions.}
\vspace{0.5cm} \noindent  
wind (Kataoka et al. 2013). Therefore, what is more surprising is that the Fermi bubbles are symmetric while the surrounding spurs are far from symmetrical, suggesting the lack of close interaction between the bubbles and Loop I in general.

These contradictions are addressed systematically by a two-step explosion process. As an initial condition, we assume that the GH was asymmetric with respect to the Galactic plane, so much so that the gas density in the northeastern area of the halo was enhanced. The temperature of the GH was almost uniform and approximated as $kT$ $\simeq$ 0.2~keV. The first explosion, either starburst activity or an AGN-like outburst, occurred in the GC approximately 15--25 Myr ago, releasing a total energy of 10$^{56-57}$ erg. The expansion velocity of the shock wave was $v_{\rm sh}$ $\sim$ 300 km s$^{-1}$ (or Mach number $M$ $\simeq$ 1.5), which slightly increased the temperature of the GH gas to $kT$ $\simeq$ 0.3~keV, forming a dense and compressed giant structure such as the NPS and Loop I. Then, approximately 5--10 Myr ago, the second explosion or energetic outflow occurred in the GC and released the energy of 10$^{55-56}$ erg. Because the first explosion had blown away most of the halo gas, the Fermi bubbles that evolved below and above the GC were almost symmetrical. 
Finally, the NPS and the northeastern bubbles were in contact owing to closer distance to the GC at the low Galactic latitude, but left a cavity between Loop I and the northwestern part of the bubble at the high Galactic latitude.
 Although speculative, further deep observations of the bubbles, NPS and Loop I with soft X-ray spectrometer (Resolve) onboard the Japanese X-ray Recovery Mission (XARM) scheduled for 2021 will enable further progress toward clarifying the past activity in the GC and its relation to the Fermi bubbles and NPS/Loop I. Specifically, measurement of the metal enrichment of the NPS/Loop I regions is crucial to reveal the physical origin of the bubbles, either past AGN-like activity or nuclear star-forming activity in the GC (e.g., Inoue et al. 2015).

\acknowledgments
Work by M.A. is supported by JSPS KAKENHI Grant Numbers JP17H06362.
 M.A. acknowledges the support from JSPS Leading Initiative for
 Excellent Young Researchers program.

\vspace{0.5cm}
\centerline{{\vbox{\epsfxsize=230pt \epsfbox{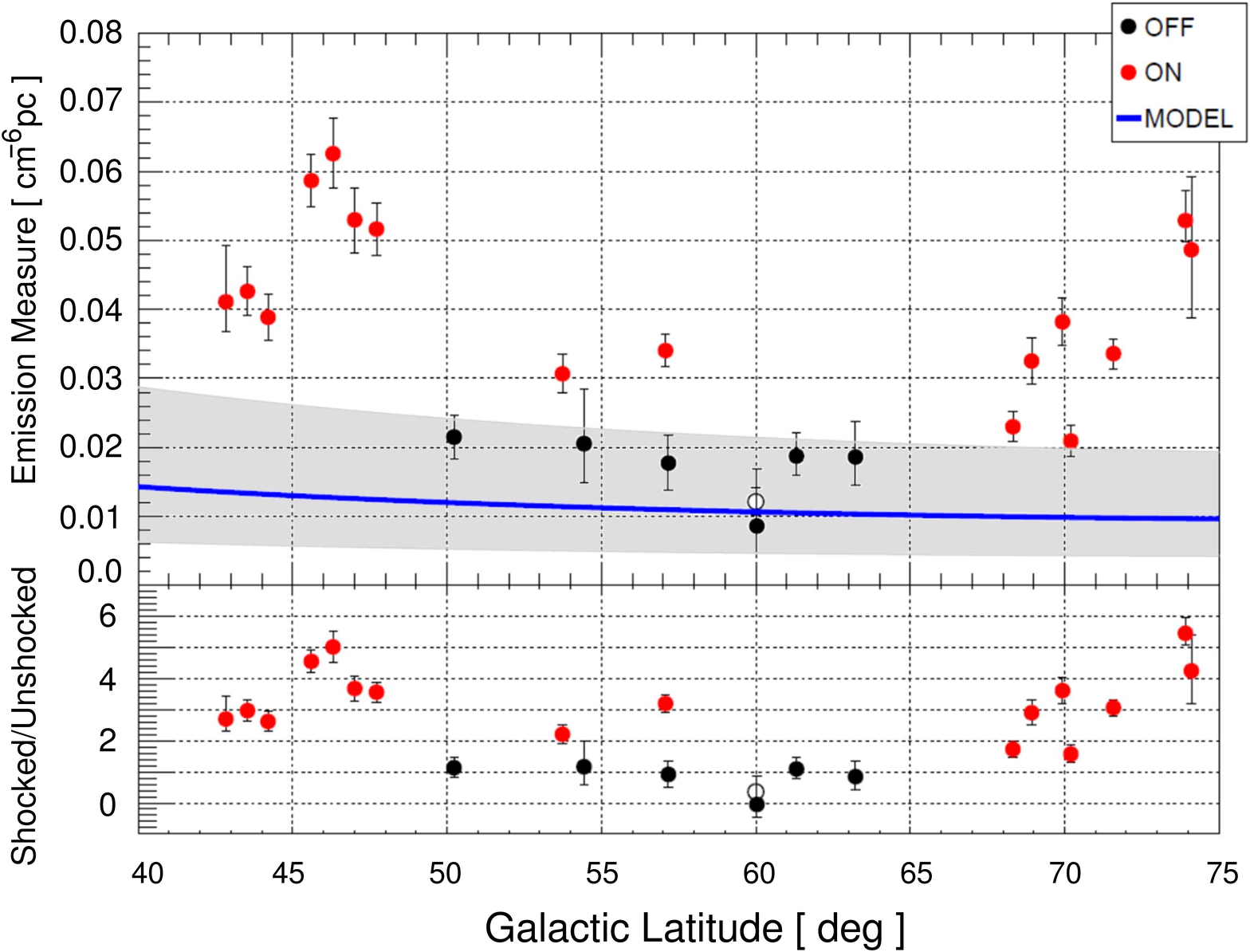}}}}
\hspace{-10truemm}
\figcaption{\small ($top$) Same as Figure.3 ($top$), but anticipated contribution of a disk-like hot gas in the GH modeled in Sakai et al. (2014) is plotted as a $thick$ $blue$ $line$ with possible uncertainties ($thin$ $grey$ $lines$). ($bottom$) the ratio of EMs in the shocked and unshocked halo gas, where the contribution of the shocked gas is simply estimated as the difference between the observed EM and the prediction of disk-like model. }
\vspace{0.5cm}

\end{document}